\documentclass[prl,twocolumn,showpacs]{revtex4}
\usepackage{graphicx}

\newcommand{\tsfrac}[2]{{\textstyle\frac{#1}{#2}}}

\newcommand{\tran}{{\scriptscriptstyle\top}}
\newcommand{\si}{\mathcal{S}}
\newcommand{\Zsi}{\mathcal{Z}_{\mathcal{S}}}

\begin{document}

\title{A Radiation Scalar for Numerical Relativity}
\author{Christopher Beetle and Lior M.~Burko}
\affiliation{Department of Physics, University of Utah, 
  Salt Lake City, Utah 84112}
\date{\today}

\begin{abstract}\noindent
This letter describes a scalar curvature invariant for general relativity with a certain, distinctive feature.  While many such invariants exist, this one vanishes in regions of space-time which can be said unambiguously to contain no gravitational radiation.  In more general regions which incontrovertibly support non-trivial radiation fields, it can be used to extract local, coordinate-independent information partially characterizing that radiation.  While a clear, physical interpretation is possible only in such radiation zones, a simple algorithm can be given to extend the definition smoothly to generic regions of space-time.
\end{abstract}

\maketitle

The prediction of the observable properties of gravitational radiation is an important application of the theory of general relativity.  There has been a recent flurry of activity in gravitational wave research as terrestrial detectors have been built to provide the first direct detection of the phenomenon.  These experiments are so delicate, however, that detailed theoretical predictions are required when generating filters to identify physical radiation in the data.  To provide these filters, researchers have turned to computers.  An impressive array of techniques have been developed which, in the near future, are likely to allow dynamical simulations of astrophysically relevant systems according to the full, non-linear Einstein equations.  Once the technical obstacles to such simulations have been overcome, however, tools will be needed to interpret their results.  Here, we propose a tool specifically designed to extract gauge-invariant information about gravitational radiation from space-time geometry.  Our proposal centers on a scalar quantity constructed from the space-time curvature which, in regions where gravitational radiation is unambiguously defined, manifestly contains information only about the radiation field.  We refer to such quantities as \textit{radiation scalars}.  While the physical interpretation of our radiation scalar is clear only in special regions of space-time, its definition (generically) extends smoothly throughout.  We present a simple algorithm to calculate it everywhere in space-time at only a modest computational expense.

The notion of gravitational radiation in general relativity does not have universal meaning.  Rather, a proper definition can be given only in regions of space-time whose geometry is characterized by two distinct length scales: one describing an ``average'' radius of curvature and a second, much shorter scale corresponding to the wavelength of gravitational waves.  Theoretical tools have been developed which allow a detailed analysis of radiation fields in such \textit{radiation zones}.  The best-known examples include the standard treatment of weak waves on flat space-time, asymptotic techniques at null infinity \cite{Arm}, and the Teukolsky formalism for weak radiation in the exterior of an equilibrium black hole \cite{Teq}.  All of these approaches separate the physical metric into a background piece and a perturbation thereof describing radiation.  They are therefore inappropriate in regions of space-time supporting stronger gravitational fields like those typically modeled in numerical relativity.

Physically, one expects a typical space-time will admit radiation zones both at large distances from any radiating sources, and also closer to them at late times as they approach a quiescent final state.  However, there are difficulties realizing this picture in numerical relativity.  Limitations imposed by computer technology make it difficult to simulate evolution at large enough distance or to late enough times to deploy perturbative techniques unambiguously.  Even within a radiation zone, it is difficult to separate a single, physical metric which has been calculated numerically into background and perturbation.  There may be many flat (or Kerr) metrics on a finite region of space-time which approximate the physical metric to a given accuracy at every point.  Employing a radiation scalar can avoid these difficulties because it (a) is calculable everywhere in space-time and (b) does not require any background metric be chosen.  On the other hand, however, the scalar we describe below does not offer as complete a description of gravitational waves in a radiation zone as the perturbative approaches mentioned above.  In particular, it vanishes even for certain non-trivial radiation fields.  However, such fields are non-generic and unlikely to be found by numerical simulations not explicitly designed to do so.  Our goal is not a formalism suited to all possible cases, but rather an approach which will be effective in \textit{generic} situations one might reasonably expect to encounter.

The use of scalar curvature invariants in numerical relativity is quite natural.  Unlike the components of a higher-rank tensor, the value of a scalar function at any point of space-time is independent of any coordinates used to evaluate it.  In a radiation zone, for example, a scalar can easily distinguish physical radiation from oscillations in the field variables induced by a peculiar choice of coordinates.

Curvature invariants have been studied extensively in the literature \cite{Wic, esb}, particularly in connection to the algebraic (Petrov) classification of space-times \cite{dIRC}.  The best-known, and algebraically simplest, examples are the two complex quantities 
\begin{equation}\label{IJ}
  \begin{array}{@{}r@{}l@{}}
    I &{}:= \tsfrac{1}{16} \left( C_{ab}{}^{cd} \, C_{cd}{}^{ab} - i \,
      C_{ab}{}^{cd} \,{}^\star\! C_{cd}{}^{ab} \right) \\[1ex]
    J &{}:= \tsfrac{1}{96} \left( 
      C_{ab}{}^{cd} \, C_{cd}{}^{ef} \, C_{ef}{}^{ab} - i \,
      C_{ab}{}^{cd} \, C_{cd}{}^{ef} \,{}^\star\! C_{ef}{}^{ab} \right), 
  \end{array}
\end{equation}
where $\,{}^\star\! C_{ab}{}^{cd} := \frac{1}{2} \epsilon_{ab}{}^{mn} \, C_{mn}{}^{cd}$ denotes the space-time Hodge dual of the Weyl tensor.  In fact, these two scalars are not only simple, but also the \textit{only} independent, algebraic curvature scalars one can construct.  This is because any curvature invariant should be expressible as a function of a set of six fundamental (real) invariants known as the eigenvalues of the Weyl tensor.  Since the traces of both the Weyl tensor and its dual vanish, a simple counting argument shows $I$ and $J$ alone suffice to reconstruct those six eigenvalues.  Thus, any scalar curvature invariant should be expressible as some function of $I$ and $J$.  The operative question then is \textit{which} such functions define radiation scalars.

Neither quantity defined in Eq.~(\ref{IJ}) is itself a radiation scalar.  The stationary Kerr black hole, for example, is known to contain no gravitational radiation anywhere in space-time.  However, neither $I$ nor $J$ vanishes in the Kerr space-time.  Intuitively, both contain contributions not just from the radiative part of the gravitational field, but also from the Coulombic part.  We must therefore seek a non-trivial function of both $I$ and $J$ in which their Coulombic contributions cancel.  One can adopt a systematic approach to this search.

The Newman--Penrose formalism \cite{NP} encodes the ten independent components of the Weyl tensor at each point of space-time in the five complex scalars 
\begin{equation}\label{Psi}
  \begin{array}{@{}r@{}l@{}}
    \Psi_0 &{}:= C_{abcd} \, \ell^a   \, m^b \, \ell^c   \, m^d \\[.5ex]
    \Psi_2 &{}:= C_{abcd} \, \ell^a   \, m^b \, \bar m^c \, n^d \\[.5ex]
    \Psi_4 &{}:= C_{abcd} \, \bar m^a \, n^b \, \bar m^c \, n^d 
  \end{array}\quad
  \begin{array}{@{}r@{}l@{}}
    \Psi_1 &{}:= C_{abcd} \, \ell^a   \, m^b \, \ell^c   \, n^d \\[.5ex]
    \Psi_3 &{}:= C_{abcd} \, \ell^a   \, n^b \, \bar m^c \, n^d 
  \end{array}
\end{equation}
These \textit{curvature components} $\Psi_n$ depend on one's choice of a null tetrad $(\ell^a, n^a, m^a, \bar m^a)$ on space-time.  All four vectors in such a tetrad are null, with the first pair real and the second both complex and conjugate to one another, and their only non-vanishing inner products are $\ell^a \, n_a = -1$ and $m^a \, \bar m_a = 1$.  The Newman--Penrose approach is particularly useful in the search for radiation scalars because it already expresses space-time curvature in the coordinate-independent scalars $\Psi_n$.  However, this independence comes at a price.  In general, there is no preferred null tetrad on space-time, and clearly different choices in Eq.~(\ref{Psi}) will lead to different values of $\Psi_n$.  There are then two ways to extract invariant, physical results.  One can either seek combinations of the $\Psi_n$ which take the same value in every frame (such as $I$ and $J$), or find a means of ``gauge-fixing'' by imposing additional conditions on the tetrad.  We shall employ a combination of these techniques below.  We should emphasize that, while the Newman--Penrose formalism is a convenient tool in our search, the final result we present below does not depend on it.  Rather, the radiation scalar we propose can be calculated directly from the space-time curvature; it is a particular function of $I$ and $J$.

In a local reference frame adapted to a given null tetrad $(\ell^a, n^a, m^a, \bar m^a)$, the $\Psi_n$ have specific physical interpretations \cite{Sgc}.  The components $\Psi_0$ and $\Psi_1$ ($\Psi_4$ and $\Psi_3$) describe transverse and longitudinal gravitational waves, respectively, propagating in the null direction $\ell^a$ ($n^a$), while $\Psi_2$ represents the Coulombic part of the field.  These interpretations are supported by results in the weak-field limits associated with radiation zones.  Moreover, both near null infinity and in the Teukolsky formalism, physical radiation is associated with the transverse components $\Psi_0$ and $\Psi_4$ while the longitudinal components $\Psi_1$ and $\Psi_3$ are understood to be ``pure gauge.''  In the Teukolsky formalism, for example, there is a natural null tetrad.  It is the \textit{Kinnersley frame} \cite{Ktd}, whose real null vectors lie in the two repeated principal null directions of the background Kerr geometry. The null tetrad in the Teukolsky formalism differs from the Kinnersley frame only at first order in perturbation theory.  However, the exact perturbation in the tetrad can always be chosen such that 
\begin{equation}\label{tf}
  \Psi_1 = 0 = \Psi_3.
\end{equation}
Motivated by these considerations, we seek to define a radiation scalar which is a function solely of $\Psi_0$ and $\Psi_4$.  However, the apparent simplicity of this goal is misleading.  It only acquires meaning if one can find a null tetrad everywhere on space-time which passes over smoothly to the correct limits at late times and large distances.

While the condition Eq.~(\ref{tf}) has been introduced in the context of perturbation theory on Kerr space-time, there is no obvious problem imposing it more generally.  We shall call a null tetrad satisfying these conditions a \textit{transverse frame}.  It is not immediately clear whether transverse frames always exist and, if they do, whether they are unique.  However, it is possible to answer both of these questions locally at any point of any space-time.

When they do exist, transverse frames are never unique.  This is because there are two types of null tetrad transformation which preserve the conditions of Eq.~(\ref{tf}).  These include a continuous family of \textit{spin-boost} transformations $S(c)$ which rescale both real null tetrad vectors and rotate the complex ones by a phase, as well as a discrete \textit{exchange} transformation $E$ which interchanges the real and complex vectors in pairs.  They affect the curvature components according to 
\begin{equation}\label{sbex}
  S(c) : \Psi_n \mapsto c^{2 - n} \, \Psi_n \quad\mbox{and}\quad E : \Psi_n \mapsto \Psi_{4 - n}.
\end{equation}
These ambiguities in a transverse frame always exist, however, and may not exhaust one's freedom in choosing it.  It is therefore natural to ask how many distinct transverse frames \textit{up to spin-boost and exchange} exist at a given point of space-time.  This question has a remarkably simple answer.  Each transverse frame is associated with a non-null, self-dual eigenvector of the Weyl tensor.  This association is not one-to-one, but rather two transverse frames give the same eigenvector if and only if they can be transformed into one another using only spin-boost and exchange operations.  While the details of the association are unimportant here, it gives the crucial result that the number of distinct transverse frames is completely determined by the algebraic (Petrov) class of the Weyl tensor.  In the various cases, one finds 
\begin{description}
\item[Type I] There are exactly three transverse frames.
\item[Type I-D] There are infinitely many transverse frames, falling in two sets.  One set contains only the Kinnersley frame, while the other contains a continuum of other transverse frames.
\item[Type II] There is only one transverse frame.
\item[Type II-N] There are a continuum of transverse frames.
\item[Type III] There are no transverse frames.
\end{description}
Geometries of types II-N and  III describe pure radiation fields, and are rarely encountered in numerical relativity unless explicitly sought.  The absence or surplus of transverse frames in these cases is therefore of little practical consequence.  Moreover, a type I-D geometry's Kinnersley frame is singularly important physically. Thus, in all cases of primary physical interest, transverse frames do exist.  The relevant ones are unique up to spin-boost and exchange, modulo an additional \textit{finite} ambiguity in the type I case which we resolve below.

In the absence of additional structures, such as a Killing field, one cannot proceed to eliminate the residual spin-boost and exchange ambiguities in a transverse frame.  We therefore adopt a different approach.  According to the transformation laws, Eq.~(\ref{sbex}), the quantities  
\begin{equation}\label{xiDef}
  \xi := \Psi_0^\tran \, \Psi_4^\tran
\end{equation}
take the same value for all transverse frames related to one another by residual transformations.  (The notation $\Psi_n^\tran$ indicates evaluation in a transverse frame.)  In a type I geometry, there will be three distinct, non-zero values of $\xi$, while for type II the only value is $\xi = 0$.  There will be two distinct values in a type I-D geometry.  One, $\xi = 0$, is associated with the Kinnersley frame, and the other with the continuum of other transverse frames.  Motivated by all these facts, we seek a function on space-time which equals one of the values of $\xi$ at every point.  If this function always approaches zero in the limits of late times and large distances, it will define a radiation scalar.

At first, the prospect of evaluating $\xi$ numerically seems daunting.  Indeed, its definition involves one or more algebraically preferred null tetrads which perhaps cannot easily be found numerically at each space-time point.  However, the close connection between transverse frames and the algebraic structure of the Weyl tensor suggests an alternative.  One can construct a tensor $\Xi_{ab}{}^{cd}$, a quadratic polynomial in $C_{ab}{}^{cd}$, whose eigenvalues are precisely $\xi$.  The characteristic equation for this tensor gives 
\begin{equation}\label{cubic}
  P := \xi^3 - \tsfrac{3}{2} I \xi^2 + \tsfrac{9}{16} I^2 \xi -  \tsfrac{1}{16} \left( I^3 - 27 J^2 \right) = 0.
\end{equation}
The three roots of $P$ are the values of $\xi$.  This result shows explicitly $\xi$ can be evaluated directly without invoking the Newman--Penrose formalism.

The roots of Eq.~(\ref{cubic}) take a particularly simple form when expressed in terms of the Baker--Campanelli \textit{speciality index} \cite{BC} $\si := 27 J^2 / I^3$.  This quantity equals unity if and only if space-time is of types II or I-D (it is ill-defined for types II-N and III, where $I = 0 = J$).  The classical Cardano formula yields
\begin{equation}\label{soln}
  \xi = I \, Z(\si) := \frac{I}{4} \left( 2 - 
    \left[ W(\si) \right]^{1/3} - \left[ W(\si) \right]^{-1/3} \right), 
\end{equation}
where $W(\si) := 2 \si - 1 + 2 \sqrt{\si^2 - \si}$.  The three values of the cube root yield the three solutions of Eq.~(\ref{cubic}).

\begin{figure}
  \includegraphics[width=\columnwidth]{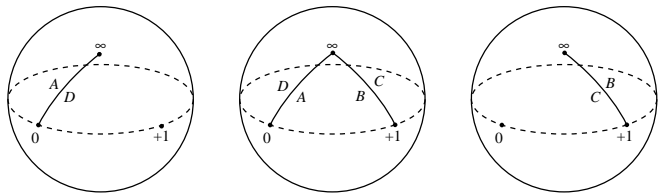}
  \caption{The Riemann surface $\Zsi$ is a triple-cover of the Riemann 
    sphere.  The values of $Z(\si)$ on its three sheets correspond to the 
    three values of $\xi$.  The labels in the figure indicate contiguous 
    regions of $\Zsi$ pictured on different sheets.}\label{Riem}
\end{figure}

The function $Z(\si)$ is not single-valued, and one must therefore pick a branch of it to evaluate $\xi$ numerically.  A careful analysis shows the Riemann surface $\Zsi$, depicted in Fig.~\ref{Riem}, for $Z(\si)$ consists of three sheets with branch points at $0$, $1$, and $\infty$.  There is a unique sheet $\Zsi^0$ with no branch point at $\si = 1$.  The associated branch $Z_0(\si)$ is unique among the three in that $Z_0(1) = 0$, whence only this branch gives the correct limit, $\xi \sim 0$, in the radiation zones at late times or large distances.  Thus,  a function which equals $\xi_0 := I \, Z_0(\si)$ in these limits will define a radiation scalar.  One can evaluate $\xi_0$ numerically using branches of the roots in Eq.~(\ref{soln}) with $\sqrt[3]{1} = \sqrt{1} = 1$, and branch cuts for both along the negative real axis.  With these choices, the branch cut for $Z_0(\si)$ itself also lies along the negative real axis.

A potential problem arises as one attempts to extend $\xi_0$ into regions of space-time supporting stronger gravitational fields.  Namely, at certain points along a curve connecting a well-understood radiation zone to the strong-field interior, the value of $\si$ may lie on the branch line for $Z_0(\si)$.  At such points, $\xi_0$ as defined above will be discontinuous.  However, although a given branch of a multi-valued complex function is necessarily discontinuous at a branch line, it is \textit{always} continuous there with one of the \textit{other} branches.  In our case, this means we can define a function along the curve in question which equals $\xi_0$ for points with values of $\si$ just to one side of the branch line, and equals one of the other roots, say $\xi_+$, for nearby points with $\si$ just to the other side.  This function is guaranteed to be smooth on space-time, and is simple enough to evaluate numerically.  Begin by calculating all three roots of Eq.~(\ref{cubic}) at each point of space-time.  At late enough times and large enough distances, pick the root $\xi_0$.  Then, extend the function globally by moving to points with successively stronger fields, always picking the appropriate root so as to keep it smooth.  This algorithm, for a generic space-time, will result in a smooth function whose large-distance and late-time limits will equal $\xi_0$.  It is a radiation scalar.

The speciality index is a convenient measure of whether a simulation has reached far enough into the asymptotic regions for the root $\xi_0$ to be clearly relevant.  Namely, if one finds in numerical data a fairly large boundary region where $|\si - 1| < 1$, it becomes reasonable to assume curves extending outward to infinity will not support values of $\si$ lying on the branch line for $Z_0(\si)$.  Practically speaking, this gives a \textit{quantitative} indication of the whether the above construction can be meaningfully applied to a given simulation.

One might worry the radiation scalar we have described is not single-valued.  Indeed, it is defined by extending its value smoothly along curves linking well-understood radiation zones to other, strong-field regions of space-time.  In principle, the value at a point in the strong-field region could depend on the curve used to reach it.  However, for the vast majority of space-times, this worry is unfounded.  Specifically, if the algebraic class of space-time is nowhere \textit{exactly} type II or higher (II-N or III; there is no problem in type I-D), the radiation scalar is necessarily single-valued.  It may be so in other situations as well, but a proof does not exist.

A conspicuous feature of the definition of $\xi$ in Eq.~(\ref{xiDef}) is its dependence on both $\Psi_0$ and $\Psi_4$.  Near infinity, these curvature components describe the ingoing and outgoing parts, respectively, of the radiation field.  There are therefore non-trivial (pure ingoing or outgoing) radiation fields for which $\xi$ will vanish.  However, such configurations are not generic.  Real radiation fields experience some back-scattering, even at large distances, and only for very carefully designed states will $\Psi_0$ or $\Psi_4$ vanish exactly.  In fact, the ingoing and outgoing parts of the radiation field are coupled through the Teukolsky--Starobinsky identities \cite{PTid, Chandra}; complete knowledge of either one determines the other.  Thus, for example, insisting the ingoing part vanishes exactly imposes severe restrictions also on the outgoing part.  As a result, pure ingoing or outgoing fields are more difficult to construct, and rarer, than one might first expect.

As a sample application of these techniques, consider a full, non-linear simulation of  a binary black hole system.  After coalescence, as the single remaining hole settles down to a quiescent final state, the relevant transverse frame will approximate the Kinnersley frame of some (undetermined) background Kerr geometry.  The factors, $\Psi_0^\tran$ and $\Psi_4^\tran$, defining $\xi_0$ will therefore satisfy the Teukolsky equation.  Their late-time behavior is known from perturbation theory: after quasi-normal ringing, both factors decay to zero as a power of time.  Moreover, the (complex) quasi-normal frequencies are specific functions of the mass and angular momentum of the final black hole.  One can therefore extract these physical parameters from the late-time behavior of the radiation scalar without first invoking perturbation theory techniques.  This is important conceptually since the ideal background for a given simulation would have the physical parameters so determined.  They would not otherwise be known when making the transition to perturbation theory.

While it may be possible to construct radiation scalars other than the one described here, they may not be easy to find.  A natural candidate, $\varrho := \si - 1$, has many features similar to those of $\xi_0$, but does not have the correct behavior near null infinity.  To leading order in perturbation theory, this quantity is given by $\psi_0^\tran \, \psi_4^\tran / (\psi_2^\tran)^2$.  While $\varrho$ certainly vanishes if space-time is exactly Kerr, the denominator here generally falls off at the same rate as the numerator.  Although the asymptotic value will be small, quadratic in the perturbation parameter, it will not fall off to zero.  It therefore does not extract information solely about radiation near null infinity.

The authors are indebted to the members of the relativity group at the University of Utah, and especially Richard Price, for many stimulating discussions.  This research has been supported by NSF grant PHY-9734871.

\end{document}